\documentclass[10pt,twoside,twocolumn,aps,pra,showpacs,
  longbibliography]{revtex4-2}
\usepackage[latin9]{inputenc}
\usepackage{color}
\usepackage{float}
\usepackage{soul}
\usepackage{tikz}
\usepackage{mathtools}
\usepackage{silence}
\usepackage{amsmath}
\usepackage{amsthm}
\usepackage{amssymb}
\usepackage{graphicx}
\usepackage{placeins}
\usepackage{afterpage, comment}
\usepackage[unicode=true,
 bookmarks=false,
 breaklinks=false,pdfborder={0 0 0},backref=false,colorlinks=true]{hyperref}\hypersetup{allcolors=blue}
 \definecolor{burgundy}{rgb}{0.5, 0.0, 0.13}
 \definecolor{amber}{rgb}{1.0, 0.75, 0.0}

\makeatletter

\usepackage{color}
\usepackage{amsthm}

\usepackage{soul}
\definecolor{cream}{rgb}{1.0, 0.99, 0.82}

\definecolor{babypink}{rgb}{0.96, 0.76, 0.76}

\usepackage{appendix}
\usepackage{epsfig}
\usepackage{newlfont}\usepackage{amsfonts}\usepackage{bm}\usepackage{subfigure}\usepackage{palatino}\usepackage{amsthm}\usepackage{braket}\usepackage{times}\usepackage{soul}\usepackage{enumitem}\usepackage{color}\usepackage[normalem]{ulem}

\makeatother
\begin{document}
\title{Characterizing errors in parameter estimation by local measurements}
\title{Characterizing errors in parameter estimation by local measurements}

\author{Riddhi Ghosh}
\email{riddhighosh22@gmail.com}
\affiliation{School of Mathematical and Physical Sciences, Macquarie University, Sydney, Australia}

\author{Alexei Gilchrist}
\affiliation{School of Mathematical and Physical Sciences, Macquarie University, Sydney, Australia}

\author{Daniel Burgarth}
\affiliation{Friedrich-Alexander-Universit\"{a}t Erlangen-N\"{u}rnberg, Germany}

\begin{abstract}
The indirect estimation of couplings in quantum dynamics relies on the measurement
of the spectrum and the overlap of eigenvectors with some reference
states. These data can be obtained by local measurements on some
sites and eliminate the need for full Hamiltonian tomography. For a 1D
chain, access to only one edge site is sufficient to compute all
the couplings between the adjacent sites, and consequently to reconstruct the full Hamiltonian.
However, its robustness in the presence of perturbations remains a critical question, particularly
when sites interact with other lattice sites beyond nearest
neighbors.

Our work studies the applicability of schemes designed for one-dimensional (1D) chains to topologies with interactions beyond nearest-neighbor. We treat interactions between the next-nearest sites as perturbation of strength $\varepsilon$ and show that the error in estimation of couplings scales linearly with $\varepsilon$ in the presence of such interactions. Further, we show that, on average, the existence of couplings between sites beyond the next-nearest-neighbor results in higher error. We also study the length of the chain that can be estimated (up to a fixed precision) as a function of $\varepsilon$, in the presence of next-nearest-neighbor interactions. Typically, for weak interactions, chains of 30 sites can be estimated
within reasonable error. Thus, we study the robustness of estimation scheme designed for a 1D chain when exposed to such multisite perturbations, offering
valuable insights into its applicability and limitations.
\end{abstract}
\maketitle

\section{Introduction}
\label{sec:intro}

Hamiltonian tomography and parameter estimation are important
for the design of quantum devices for quantum technologies
such as quantum computation, communication, and metrology. Since these quantum devices need to be carefully isolated to prevent unwanted interactions and noise, they are often difficult to probe, and characterization methods are frequently constrained. Various methods for estimating the parameters of the Hamiltonian governing the system have been explored.

Such characterization schemes span from performing full informationally complete tomography to the application of machine learning algorithms on restricted data. Several studies leverage maximum likelihood estimation for deducing the Hamiltonians based on local measurements~\cite{ref_1,ref_2,ref_3,ref_4,ref_5,ref_6,ref_7}. Notably, maximum likelihood estimation can be merged with matrix product operators to provide a scalable estimation method~\cite{ref_ex_1}. 
Methods for quantum-state or process tomography using linear regression methods have also been studied~\cite{DDong1,DDong2}. In Refs.~\cite{ref_ml_1,ref_ml_2,ref_ml_3}, the authors present an algorithm for Hamiltonian learning using copies of the Gibbs state. Other approaches involve optimal measurement strategies for extracting maximum information from measurement and methods from sensing and metrology~\cite{ref_sen_1,ref_sen_2,ref_sen_3,ref_sen_4}.

A notable approach proposed in Ref.~\cite{DB1} is the method of indirect estimation of couplings through local measurements conducted on a subset of the sites within a specified lattice. This method relies on the precise determination of the spectrum and the overlap of the eigenvectors with designated reference states; these represent local probes. For example, in a one-dimensional system with $N$ sites, complete knowledge of the spectrum $\left\{ e_{k}\right\}$, and the overlap between each eigenvector and a reference state, say $|1\rangle$ corresponding to a local probe measured at one end of the chain (quantified by $\left|\left\langle 1|e_{k}\right\rangle \right|^{2}$) is sufficient to estimate all the couplings between the sites. While the spectrum constitutes a global property of the entire chain,
it can be effectively extracted from the free-induction decay (FID) signal obtained by initializing and measuring a single excitation at one end of the chain~\cite{DB1}. Specifically, the Fourier transform of the FID reveals both the energy eigenvalues and the squared overlaps $\left|\left\langle 1|e_{k}\right\rangle \right|^{2}$, thereby providing the required spectral information through local measurements. Alternatively, one may measure only the spectra, but in the presence of a local perturbation~\cite{DB2}. For a one-dimensional chain with only nearest-neighbor interaction, performing
measurement at one end of the chain is sufficient; this will be discussed in more detail in Sec.~\ref{sec:frame}. This approach significantly reduces the resources that would otherwise be needed for complete state tomography, but potentially comes at a price of lower precision.

The robustness of this scheme has been studied in the context of a one-dimensional linear topology. The impact of errors occurring during measurement and experimental
implementation, encompassing factors such as insufficient resolution,
constrained signal-to-noise ratio, and various other experimental
and instrumental inaccuracies, has been explored in Ref.~\cite{Feng}. The study examines the effect of imprecise measurements that lead to partial loss of information due to broadening of the spectral peaks. To address this, the recursive relations introduced in Refs.~\cite{DB1,DB2} were modified to add a correction term designed to accommodate missing or insufficient experimental data ~\cite{Feng}. The scheme has been shown to be robust when this correction term is introduced.

Subsequent investigations have expanded on the original work in Ref.~\cite{DB1}, demonstrating the applicability of the algorithm to arbitrary networks, where measurements are carried out at multiple local sites~\cite{DB3}. Estimation of couplings for networks with interactions beyond the nearest-neighbor involves the concept zero forcing~\cite{DB3,DB4,ZF}. Zero forcing establishes a set of rules that are useful to identify the set of vertices that are sufficient for indirect estimation of couplings between all the sites. This will be discussed in detail in Sec.~\ref{sec:frame}. It is noteworthy that the concept of zero forcing is intimately related with the controllability of these networks~\cite{DB4}.

Experimentally, the interplay between estimation and controllability has been exploited for quantum-state tomography using nitrogen-vacancy centers in diamond coupled to a nuclear spin in Ref.~\cite{exp1}. The study shows that when a system is controllable by a driven single qubit, then a local random pulse is sufficient for a complete quantum-state tomography. Another experimental implementations of estimation despite restricted access to the system is presented in Ref.~\cite{exp2}, where quantum system identification is achieved by analyzing free induction decay signals. The experiment is implemented using liquid nuclear magnetic resonance quantum information processor and is based on the measurement of time traces of a restricted subsystem~\cite{exp2-theo}.

It is important to acknowledge that a single site may exhibit interactions with multiple sites that extend beyond its immediate neighbors. Indeed, in the usual experimental setup, most sites within the system interact with each other, typically with a decreasing interaction strength as the distance between them increases. Sometimes, interactions beyond nearest-neighbor are sufficiently weak and can be ignored. In this paper, we study the robustness of the protocol in experimental
configurations where such long-range interactions might be relevant. Our primary focus centers on a scenario in which the presence of such couplings beyond nearest-neighbors is unknown. Consequently, we implement the protocol with the assumption of nearest-neighbor interactions and seek to determine whether the resulting errors have any significant impact on the estimation of the coupling strength. 

In Sec.~\ref{sec:frame}, we start with a description of the scheme
employed for the indirect estimation of couplings for nearest-neighbor
interactions. We provide a concise overview of the zero-forcing conditions~\cite{DB4,ZF}, which serve as the foundation for understanding how the original scheme could be adapted when interactions are known to extend beyond nearest-neighbor. Furthermore,
we present an extension of the estimation protocol tailored to accommodate
next-nearest-neighbor couplings. Our results are organized into three
distinct subsections. In Sec.~\ref{sec:reA}, we derive an analytical
error bound for estimation up to the second coupling and formulate an ansatz capturing the behavior for $n\geq3$, where the $n$th term represents coupling between $n$ and $n+1$ sites. This is followed by Sec.~\ref{sec:reB},
where we investigate the critical length, i.e., the number of couplings
that can be estimated up to a fixed precision, as a function of the strength of perturbation.
Finally, in Sec.~\ref{sec:reC}, we expand the study to systems
with interactions beyond the next-nearest-neighbors. 
\par
\section{Mathematical framework}
\label{sec:frame}

Zero forcing is a concept closely linked to the controllability of
a system with a known topology. The connection between zero forcing
and quantum system identification and parameter estimation has been
well established in previous research~\cite{DB5}. Consider a simple graph where the vertices may be white or blue, and an iterative process where a blue vertex in a graph turns a neighbor blue if and only if it is the only white neighbor of that vertex; the blue vertices remain blue. The iterative process is continued until no further change happens. A zero forcing set is a subset of vertices in a graph that turns all the other vertices blue. 

In the context of tight-binding models or spin chains, within first excitation sector, the system Hamiltonian can be represented as a hopping Hamiltonian on a graph, where the sites correspond to the vertices and the couplings to the edges. The size of a minimum zero forcing set
of a given graph is called the zero forcing number and is an upper
bound for the degeneracy of the Hamiltonian governing the system~\cite{ZF2}.
The minimum zero forcing set also gives the number of sites that are
sufficient to perform local measurements on, in order to estimate
the couplings in a given network by applying the estimation scheme~\cite{DB1,DB2}. 

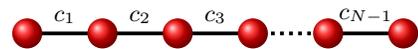
\begin{figure}[htbp]
\begin{tikzpicture}
        \foreach \i in {1,...,3}{
            \draw[line width=0.5mm,black] (\i,0) -- node[above,black] {$c_{\i}$} (\i+1,0);
        }
        \draw[line width=0.5mm,black,dotted] (4,0) -- (5,0);
        \draw[line width=0.5mm, black] (5,0) -- node[above,black] {$c_{N-1}$} (6,0);
        \foreach \i in {1,...,6} {
            \shade[ball color=red] (\i,0) circle (0.2);
        }
    \end{tikzpicture}
\caption{Topology corresponding to the Hamiltonian with nearest-neighbor interactions.
The black lines denote the couplings.}
\label{fig:chain1} 
\end{figure}

A simple example illustrating the applicability of zero forcing is that of a one-dimensional chain with $N$ sites. 
We consider a Hilbert space $\mathcal{H}$ of dimension $N$
equipped with the orthonormal basis vectors $\{|1\rangle,|2\rangle,\dots,|N\rangle\}$. A hopping Hamiltonian within this space can be written as $H_{C}=\sum_{n=1}^{N-1}c_{n}(|n\rangle\langle n+1|+|n+1\rangle\langle n|)$, where the coupling strength between the $n$th and $\left(n+1\right)$th sites is $c_{n}$, $1\leq n\leq N-1$ (see Fig.~\ref{fig:chain1}). Here, we choose $c_n$ as real and positive. This choice incurs no loss of generality in the one-dimensional case. Any complex value $c_{n}'=|c_{n}|e^{i\theta_n}$ can be mapped to a real, positive value by a local gauge transformation, specifically by applying site-dependent phase shifts $|n^{'}\rangle\rightarrow e^{i\theta_n} |n\rangle$. Such transformations do not affect the underlying physics and any complex phase can
be absorbed in a choice of basis~\cite{DB2}. 
\par
Knowing the spectrum of $H_{C}$ given by $\left\{ e_{k}\right\}$ and
the overlap of the first site with a reference state, i.e., $\langle1|e_{k}\rangle$,
we can determine that $c_{1}^{2}=\sum_{k=1}^{N}\left|e_{k}\langle1|e_{k}\rangle\right|^{2}$
and $\langle2|e_{k}\rangle={e_{k}\langle1|e_{k}\rangle}/{c_{1}}$. The required spectral data can be obtained from local dynamical measurements. If an excitation is initialized on site $\ket{1}$, the return amplitude 
\begin{equation}
    f(t) = \bra{1}e^{-iHt}\ket{1}
\end{equation}
can be recorded as a function of time. Its Fourier transform yields
\begin{equation}
    F(\omega) = \sum_k |\langle1|e_k^2\rangle|\delta(\omega - e_k),
\end{equation}
where the position of the peaks correspond to the eigenvalues $e_k$, while the heights of the peaks give the amplitudes $|\langle1|e_k^2\rangle|^2$. Thus, both the spectrum and the local projections of eigenvectors can be extracted from the FID signal measured at a single site.

The coupling constants for $n\geq2$ can be calculated using the following
recursive formulas~\cite{DB1} (see appendix~\ref{app:A} for derivation):

\begin{equation}
    c_{n}=\sqrt{\sum_{k=1}^{N}\left|e_{k}\langle n|e_{k}\rangle-c_{n-1}\langle n-1|e_{k}\rangle\right|^{2}},
\end{equation}

\begin{equation}
    \langle n+1|e_{k}\rangle=\frac{e_{k}\langle n|e_{k}\rangle-c_{n-1}\langle n-1|e_{k}\rangle}{c_{n}}.
\end{equation}

Zero forcing also indicates that in the presence of next-nearest-neighbor interactions, local measurements on more than one site may be necessary to obtain all the couplings.

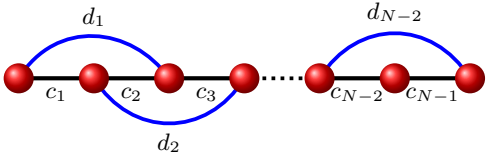
\begin{figure}[htbp]
\begin{tikzpicture}
        \foreach \i in {1,...,3}{
            \draw[line width=0.5mm,black] (\i,0) -- node[below,black] {$c_{\i}$} (\i+1,0);
        }
        \draw[line width=0.5mm, black] (5,0) -- node[below,black] {$c_{N-2}$} (6,0);
        \draw[line width=0.5mm, black] (6,0) -- node[below,black] {$c_{N-1}$} (7,0);
        \foreach \i in {1,...,3,5,6}{
            \draw[line width=0.5mm,black] (\i,0) -- (\i+1,0);
        }
        \draw[line width=0.5mm,black,dotted] (4,0) -- (5,0);
        
        \draw[line width=0.5mm,blue] (1,0) .. controls (1.5, 0.8) and (2.5, 0.7) .. node[above,black]{$d_{1}$} (3,0);
        
        \draw[line width=0.5mm,blue] (2,0) .. controls (2.5, -0.8) and (3.5, -0.8) .. node[below,black]{$d_{2}$} (4,0);
        \draw[line width=0.5mm,blue] (5,0) .. controls (5.5, 0.8) and (6.5, 0.8) .. node[above,black]{$d_{N-2}$} (7,0);
        \foreach \i in {1,...,7} {
            \shade[ball color=red] (\i,0) circle (0.2);
        }  
    \end{tikzpicture}
\caption{Topology corresponding to the Hamiltonian with nearest and next-nearest-neighbor interactions.
The black lines denote the nearest-neighbor couplings and the blue
lines denote the next-nearest-neighbor couplings.}
\label{fig:chain2} 
\end{figure}

We now treat the next-nearest-neighbor term as a perturbation to the original Hamiltonian $H_{C}$.  We use $d_{l}$ to represent the coupling between the $l$th and $\left(l+2\right)$th site ($1\leq l\leq N-2$); in accordance with $c_{n}$ and for the sake of simplicity, we choose $d_{l}$ also to be real and positive. The magnitudes of both  $c_{n}$ and $d_{l}$ are of the same order. The interaction strength of the nearest-neighbor sites is on the order of $c_{n}$ while that of next-nearest-neighbor is controlled by the strength of perturbation $\varepsilon$; i.e., the coupling is proportional to $\epsilon d_{l}$. The modified Hamiltonian is $H^{\epsilon}=H_{C}+\epsilon H_{D}$, where $H_{D}=\sum_{l=1}^{N-2}d_{l}\left(|l\rangle\langle l+2|+|l+2\rangle\langle l|\right)$
and $0\leq\epsilon\leq1$ is the strength of perturbation. The topology
corresponding to $H^{\epsilon}$ is shown in Fig.~\ref{fig:chain2}.

The spectrum of the Hamiltonian $H^{\epsilon}$ is denoted as $\left\{ e_{k}^{\epsilon}\right\} $
and the eigenvectors are $\left\{ |e_{k}^{\epsilon}\rangle\right\} $,
$k=1,\dots,N$. Building upon the calculations in the previous case,
it becomes evident that $c_{1}=\sum_{k=1}^{N}e_{k}^{\epsilon}\langle1|e_{k}^{\epsilon}\rangle\langle e_{k}^{\epsilon}|2\rangle$.
Thus, in order to compute $c_{1}$ using this estimation scheme, it
is sufficient to have the knowledge of $\left\{ e_{k}^{\epsilon}\right\} $
and the coefficients $\langle1|e_{k}^{\epsilon}\rangle$ and $\langle2|e_{k}^{\epsilon}\rangle$.
Furthermore, we observe that $\epsilon^{2}d_{1}^{2}=\sum_{k=1}^{N}\left|e_{k}^{\epsilon}\langle1|e_{k}^{\epsilon}\rangle-c_{1}\langle2|e_{k}^{\epsilon}\rangle\right|^{2}$
and $\langle3|e_{k}^{\epsilon}\rangle=({e_{k}^{\epsilon}\langle1|e_{k}^{\epsilon}\rangle-c_{1}\langle2|e_{k}^{\epsilon}\rangle})/{\epsilon d_{1}}$.
By following this methodology, it is possible to compute all the coupling
coefficients $c_{n}$'s and $d_{n}$'s for $n\geq2$ using the recursive
relation stated below:

\begin{equation}
    c_{n}=\sum_{k=1}^{N}e_{k}^{\epsilon}\langle n|e_{k}^{\epsilon}\rangle\langle e_{k}^{\epsilon}|n+1\rangle,
\end{equation}
\begin{equation}
\epsilon^{2}d_{n}^{2}=\sum_{k=1}^{N}\left|e_{k}^{\epsilon}\langle n|e_{k}^{\epsilon}\rangle-c_{n-1}\langle n-1|e_{k}^{\epsilon}\rangle-c_{n}\langle n+1|e_{k}^{\epsilon}\rangle\right|^{2},
\end{equation}
\begin{equation}
    \langle n+2|e_{k}^{\epsilon}\rangle=\frac{e_{k}^{\epsilon}\langle n|e_{k}^{\epsilon}\rangle-c_{n-1}\langle n-1|e_{k}^{\epsilon}\rangle-c_{n}\langle n+1|e_{k}^{\epsilon}\rangle}{\epsilon d_{n}}.
\end{equation}

The spectral decomposition of the dynamics of the system can be accessed via the Fourier transform of the FID, which provides the magnitudes of overlaps $|\langle1|e_{k}^{\epsilon}\rangle|^{2}$ and $|\langle2|e_{k}^{\epsilon}\rangle|^{2}$ onto the energy eigenstates~\cite{DB2}. While this provides the modulus of the overlaps, full reconstruction of the complex coefficients $\langle1|e_{k}^{\epsilon}\rangle$ and $\langle2|e_{k}^{\epsilon}\rangle$ requires knowledge of their relative phases. These phases are not accessible through FID alone but can be inferred from coherent transport measurements. As demonstrated in Ref.~\cite{DB3}, by analyzing the time resolved propagation of local excitations, one can extract the phase information necessary to reconstruct the full complex structure of the eigenstate projections. Consequently, we establish that for the topological network described by the Hamiltonian $H^{\epsilon}$, measuring two specific sites is sufficient to obtain information regarding all the couplings and therefore the Hamiltonian. This observation aligns with the zero-forcing condition.

However, if one is unaware of the next-nearest-neighbor couplings and the estimation protocol designed for nearest-neighbor interactions is applied regardless, the extended interactions effectively introduce an error in the estimated values of the couplings. This is the setup that we consider now. 

\section{Results}

\label{sec:re}

In the previous Sec., we realized that for a chain with next-nearest-neighbor interaction, the values of $\left\{ e_{k}^{\epsilon}\right\} $,
$\langle1|e_{k}\rangle$, and $\langle2|e_{k}\rangle$ are sufficient
in order to effectively deduce all the coupling constants. In the
instances where the knowledge of $\langle2|e_{k}\rangle$ is lacking or where one is unaware of next-nearest-neighbor couplings, one can still try to compute the nearest-neighbor couplings by applying the estimation scheme for a chain with nearest-neighbor interaction only. However, the coupling constants estimated using this methodology may exhibit some degree of deviation from their true values. Our aim is to find out how the magnitude of this discrepancy depends on the strength of the perturbation $\epsilon$.

\subsection{Upper bound of error in estimation for up to $N=30$}

\label{sec:reA}

Considering that we employ the estimation scheme for nearest-neighbor
coupling although our system is characterized by next-nearest-neighbor
interaction, the calculated coupling coefficients exhibit inaccuracies;
we denote these couplings as $c_{n}^{\epsilon}$. In the absence of
perturbation, i.e., when $\epsilon=0$, $c_{n}^{\epsilon}=c_{n}$.
We quantify the error at site $n$, resulting from the omission of
next-nearest-neighbor couplings as $\Delta_{n}=\sqrt{\left|\left(c_{n}^{\epsilon}\right)^{2}-c_{n}^{2}\right|}$. 

Using the formulas presented in Sec.~\ref{sec:frame}, we can derive
the expression $\left(c_{1}^{\epsilon}\right)^{2}=\sum_{k=1}^{N}\left(e_{k}^{\epsilon}\right)^{2}\left|\langle1|e_{k}^{\epsilon}\rangle\right|^{2}$
(see Appendix~\ref{app:A}). Furthermore, $\langle1|\left(H^{\epsilon}\right)^{2}|1\rangle=\sum_{k=1}^{N}\left(e_{k}^{\epsilon}\right)^{2}\left|\langle1|e_{k}^{\epsilon}\rangle\right|^{2}=c_{1}^{2}+\epsilon^{2}d_{1}^{2}$.
Consequently, the square of the error in calculating the coupling
constant between the first and second sites is (see Appendix~\ref{app:B})

\begin{equation}
    \Delta_{1}=\sqrt{\left|\left(c_{1}^{\epsilon}\right)^{2}-c_{1}^{2}\right|}=\epsilon d_{1},
\end{equation}
as is expected from perturbation theory. Now, we use the
recursive relations for nearest-neighbor interaction outlined in Sec.~\ref{sec:frame} to define the following terms, for $n\geq2$.

\begin{equation}
    \left(c_{n}^{\epsilon}\right)^{2}\coloneqq\sum_{k=1}^{N}\left|e_{k}^{\epsilon}\langle n|e_{k}\rangle_{\varepsilon}-c_{n-1}^{\epsilon}\langle n-1|e_{k}\rangle_{\varepsilon}\right|^{2},
\end{equation}
\begin{equation}
    \langle n+1|e_{k}\rangle_{\varepsilon}\coloneqq\frac{e_{k}^{\epsilon}\langle n|e_{k}\rangle_{\varepsilon}-c_{n-1}^{\epsilon}\langle n-1|e_{k}\rangle_{\varepsilon}}{c_{n}^{\epsilon}}.
\end{equation}

It is essential to note that the subscript $\varepsilon$ signifies
that the term $\langle n|e_{k}\rangle_{\varepsilon}$ is not the overlap
of $\left|n\right\rangle $ with the eigenvectors of the system, i.e.,
$\langle n|e_{k}^{\epsilon}\rangle\ne\langle n|e_{k}\rangle_{\varepsilon}$. Rather, it is the \emph{estimated} overlap -- the numerical value that we calculate from the estimation protocol, including an approximation error. We obtain
\begin{equation}
    \Delta_{2}=\epsilon\left(d_{2}+\frac{d_{1}c_{3}}{c_{1}}\right)+\mathcal{O}(\epsilon^{2})
\end{equation}

\begin{figure}[htbp]
\centering \includegraphics[width=1\linewidth]{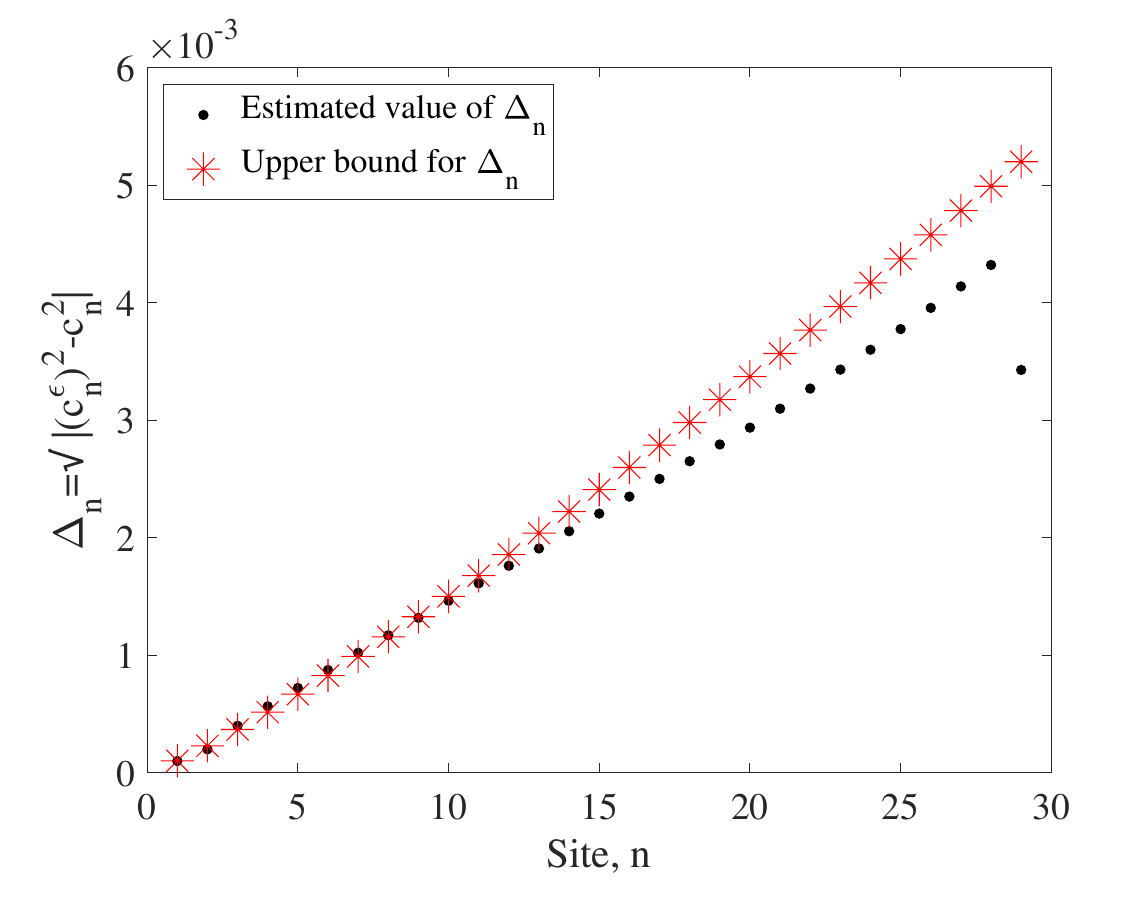} \caption{The error $\Delta_{n}$ in estimation of the coupling constants in
the presence of next-nearest-neighbor interaction. Each $\Delta_{n}$ is the average over $1000$ randomly chosen Hamiltonians; the Hamiltonians are constructed with $c_{n}$ and $d_{n}$ randomly chosen from a uniform distribution lying in $\left[0.95,1.05\right]$. The upper bound is given by $n^{\frac{7} {6}}\epsilon\left(\max\left(d_{k}\right)\right)$, which is valid for up to $N=30$. Here, $\epsilon=10^{-4}$. Note: The last point ($n=N-1$) shows a small drop due to a boundary effect in the recursion.}
\label{fig:P1} 
\end{figure}

Since $\epsilon$ is small, we ignore the higher-order terms from
$\mathcal{O}(\epsilon^{2})$ and conclude that $\Delta_{2}$ also
scales with $\epsilon$. 
Since the additional couplings are treated as perturbation, one expects errors on the order of $\epsilon$. While this scaling is evident for the first two terms ($\Delta_1$ and $\Delta_2$), obtaining closed-form expressions for higher $n$ becomes analytically intractable. The relevant quantity is not just the order of the error, but the coefficient in front of $\epsilon$, and in particular how it scales with the site $n$. We therefore introduce a heuristic ansatz to approximate the overall growth of the error with $n$, 
\begin{equation}
    \Delta_{n}\le n^{p}\epsilon\left(\max\left(d_{k}\right)\right)
    \label{eq:bound}
\end{equation}
where $p$ is an empirically chosen exponent that provides the best fit to the numerical data. This ansatz is not derived analytically but was found to accurately describe the observed scaling. A numerical study shows that for up to $N=30$, we can choose $p={7}/{6}$
to get a fairly tight upper bound for the error in estimation, as shown in Fig.~\ref{fig:P1}. It serves as a convenient empirical upper bound consistent with the linear $\epsilon$ dependence obtained from perturbation theory.

Figure~\ref{fig:P1} illustrates the behavior of estimation error when the ideal nearest-neighbor model is perturbed by weak next-nearest-neighbor couplings. The black points show the average deviation from the actual coupling strength, while the red points indicate the empirical upper bound given in Eq.~\eqref{eq:bound}. The gradual increase in the error reflects the cumulative propagation of perturbative effects through the recursive estimation scheme. Importantly, the error remains small, demonstrating the robustness of the method.

For each simulation, the $c_n$'s and $d_n$'s were chosen independently from uniform distributions centered at unity with a relative variation of $2.5\%$. Therefore, both $c_n,d_n\in\left[ 0.95, 1.05 \right]$. The result shown in Fig.~\ref{fig:P1} represent averages over $1000$ independent random realizations of these parameters.

It should also be noted here that the small drop observed in the error at the last site  (in Fig.~\ref{fig:P1}) arises from a finite-size boundary effect in the recursion procedure. At the edge of the chain, only one coupling is to be estimated at instead of two; this might make the error slightly lower. Increasing the chain size confirms this interpretation --- for example, when $N=40$, the site corresponding to $n=29$ no longer shows a drop.

\subsection{Critical length}

\label{sec:reB}

\begin{figure}[htbp]
\begin{centering}
\includegraphics[width=1\linewidth]{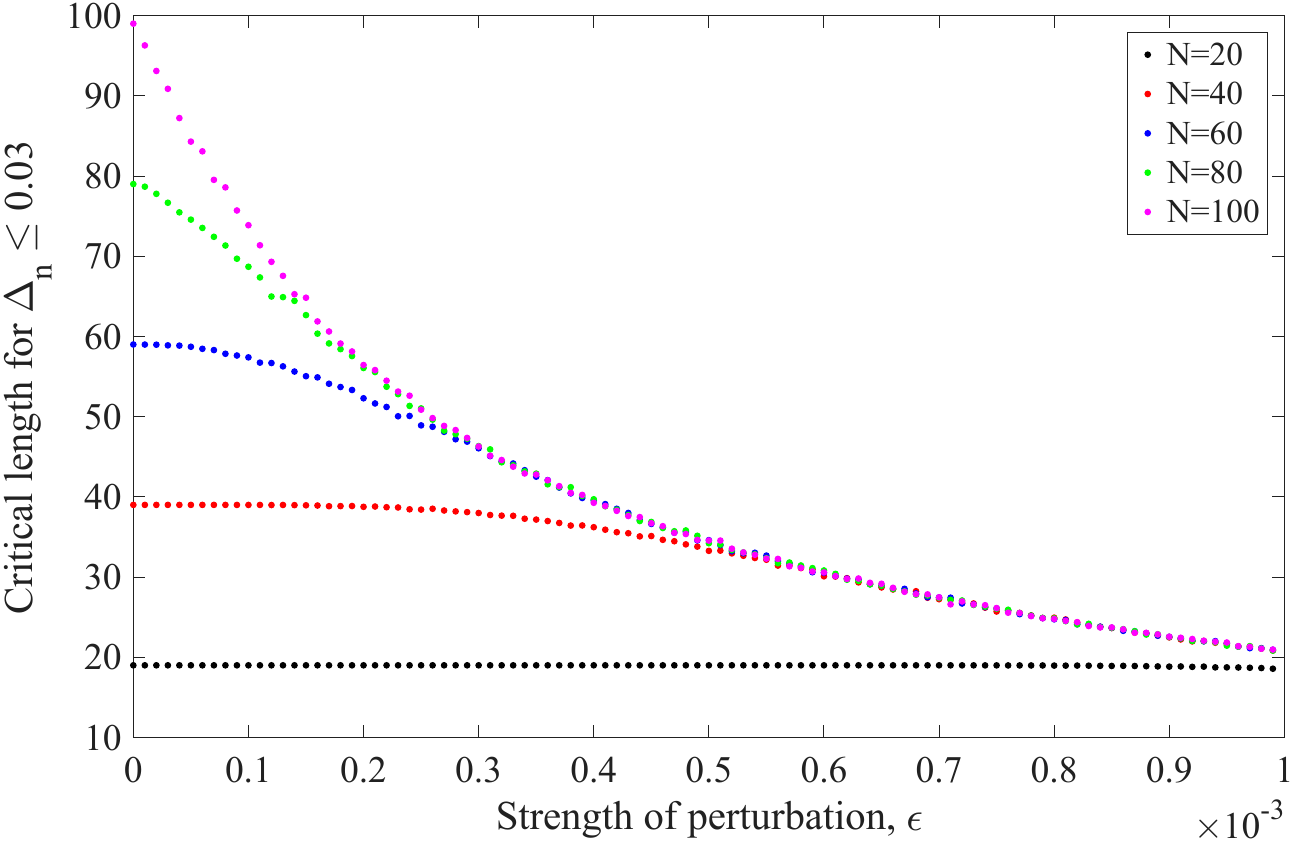}\\ (a) \\
\includegraphics[width=1\linewidth]{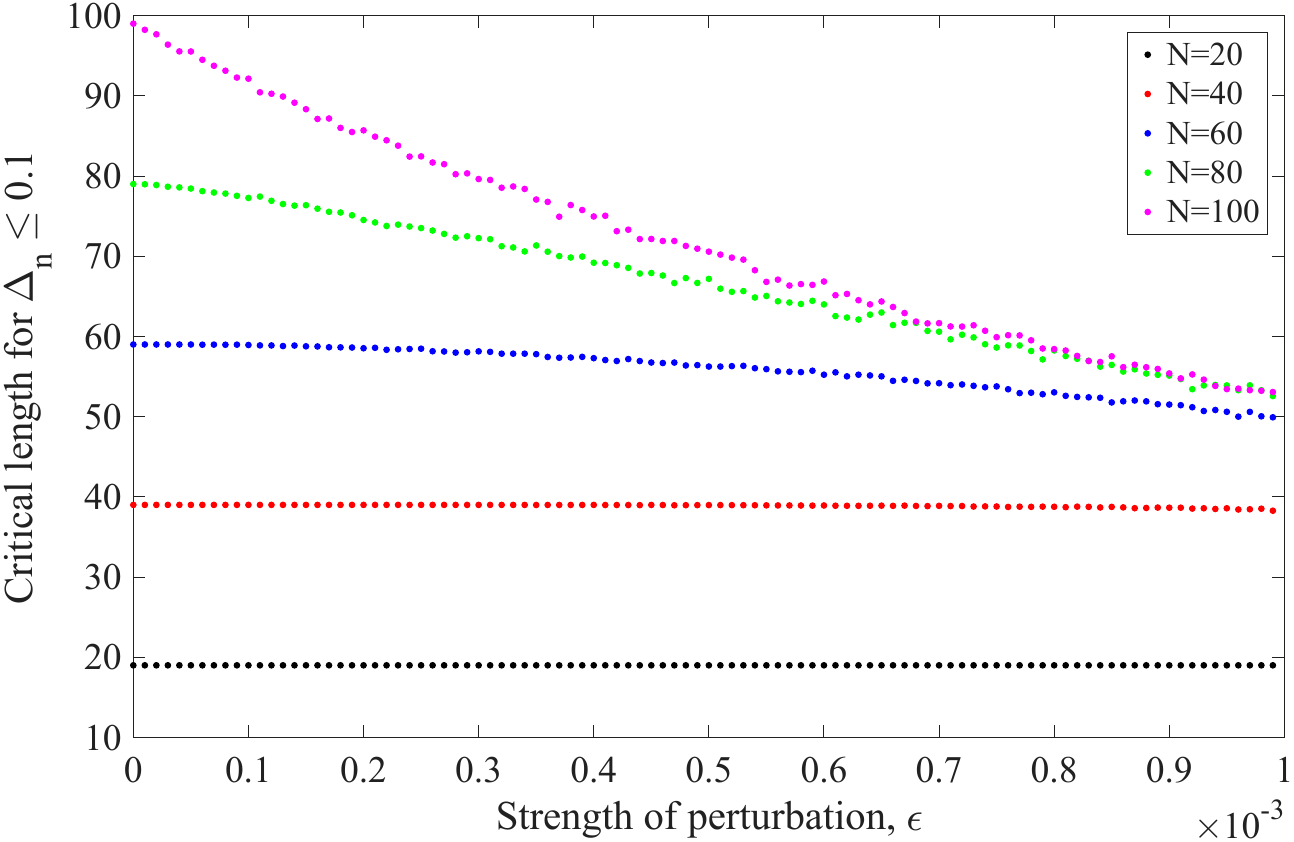}\\ (b) 
\par\end{centering}
\caption{The variation of the critical length $L_{C}$ with the strength of
perturbation $\epsilon$, up to a precision of (a) $0.03$ and (b) $0.1$, for different
lengths of chain $N$. The Hamiltonians are constructed with
$c_{n}$ and $d_{n}$ randomly chosen from a uniform
distribution lying in $\left[0.95,1.05\right]$. For $\epsilon=0$,
the critical length $L_{C}=N$; all the couplings can be estimated for any $N$ when $\varepsilon=0$.}
\label{fig:P5} 
\end{figure}

It is evident from Fig.~\ref{fig:P1} that the error in the estimated
value increases along the chain. Couplings can be estimated with reasonable
accuracy up to a certain length that depends on the value of $\epsilon$.
In this context, we introduce the concept of \textit{critical length},
$L_{C}$, which characterizes the segment of the chain (quantified
by the number of coupling constants) that can be estimated within
a predetermined margin of error, for a chain of length $N$. The behavior of $L_{C}$ is extracted from exact numerical simulations. 
Figure~\ref{fig:P5} shows the variation of $L_{C}$
with strength of perturbation $\epsilon$, revealing a clear trend -- as $\epsilon$ increases, the portion of the chain that remains accurately estimable decreases. By construction, in the absence of perturbation ($\epsilon = 0$), the estimation protocol exactly recovers all coupling, i.e., $L_C=N$, irrespective of the chain length.
 
The feasibility of the estimation algorithm becomes evident when applied
to scenarios involving relatively minor perturbations, typically up
to the magnitude of $10^{-3}$. Many experimental setups feature considerably
fewer than 50 sites. For instance, in the study presented in Ref.~\cite{IonEx},
a configuration consisting of 42 trapped ions is employed to investigate
system dynamics with local observations. The estimation algorithm should perform well with such system
sizes.

\subsection{Generalization to interactions beyond next-nearest-neighbor}

\label{sec:reC}

We now consider scenarios where the system exhibits interactions beyond nearest-neighbor. For a chain size of $N=20$, we choose two distinct graphs where the interactions occur between $18$ specific pairs of sites and compare the estimation errors with the case where next-nearest-neighbor interactions are present. In each case, the system also includes nearest-neighbor couplings. The strength of perturbation is fixed at $\varepsilon=10^{-4}$ in each of these three different cases.

\begin{figure}[htbp]
\begin{centering}
    \includegraphics[width=\linewidth]{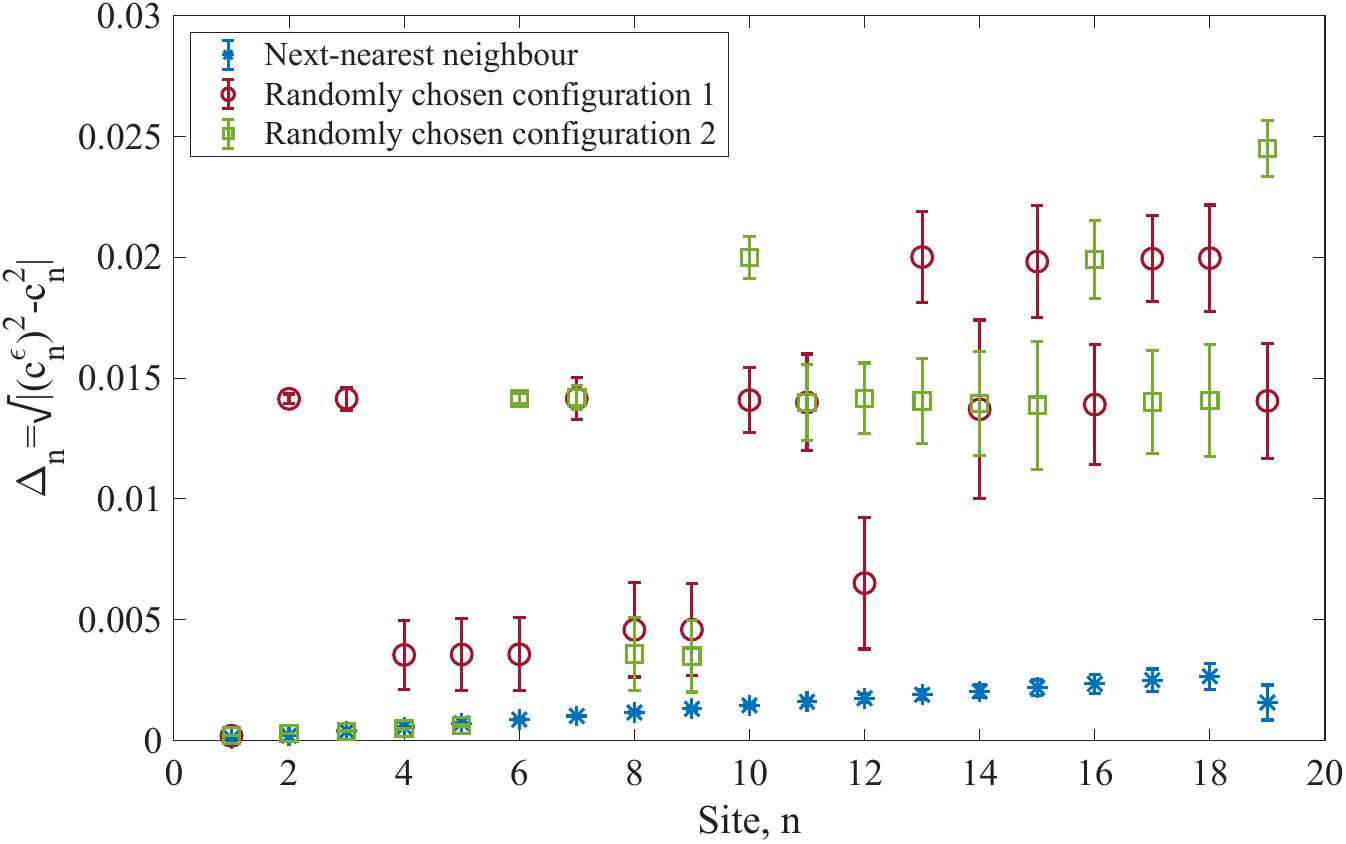}\\ (a) \\
    \includegraphics[width=\linewidth]{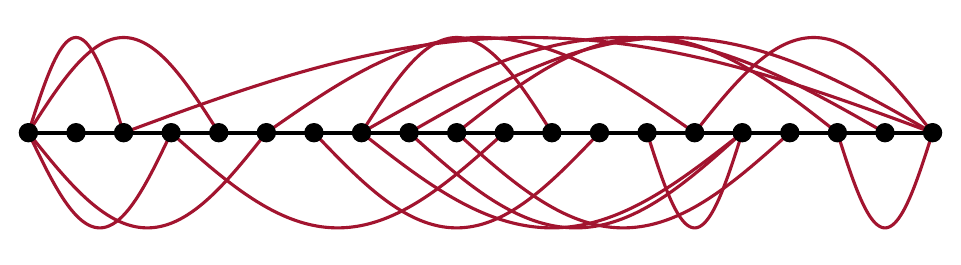}\\ (b) \\ 
    \includegraphics[width=\linewidth]{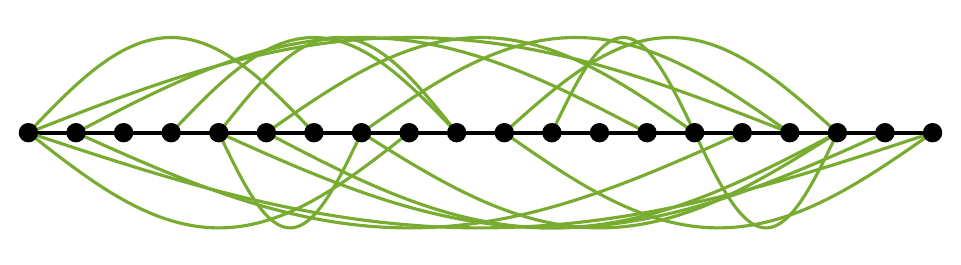}\\ (c)
\par\end{centering}
    \caption{(a) The average estimation error as a function of site $n$ for three different adjacency profiles, for a chain of size $N=20$. Each point is an average of $1000$ instances of the Hamiltonian. The blue points correspond to the next nearest-neighbor interactions. The error bars denote the standard deviation over all the random realizations. The overlap between green and blue points is expected, as these configurations yield similar average errors. The interaction structures of the chains corresponding to the red and green points, respectively, are shown in panels (b) and (c). Both cases include long-range interactions beyond next-nearest-neighbor.}
    \label{fig:Pnew1}
\end{figure}

For each graph, we generate $1000$ random instances of coupling strengths -- $c_{n}$ and $d_{n}$, sampled independently for each realization. The estimation error is then averaged over these $1000$ cases to obtain an overall trend. Figure~\ref{fig:Pnew1} shows the average estimation error for the three cases, along with the corresponding graphs.
Here, we examine the effect of perturbations arising from non-nearest-neighbor couplings. The different sets of points correspond to configurations with the same number of additional couplings but at increasing spatial separations. This approach isolates the influence of interaction distance on estimation accuracy. Simply extending the interaction range would introduce only a few such links in a finite chain, providing insufficient statistics for meaningful averaging. By keeping the number of added couplings constant while varying their distance, we can compare the impact of perturbations on equal footing across different topologies.

On average, the system with next-nearest-neighbor interactions consistently exhibit lower error. While individual realizations in the other two cases can occasionally outperform the next-nearest-neighbor setup, the overall trend is clear. Additionally, the standard deviation in error is significantly smaller in the next-nearest-neighbor case, indicating more robust and predictable estimation performance.

\begin{figure}[H]
\begin{centering}
\includegraphics[width=1\linewidth]{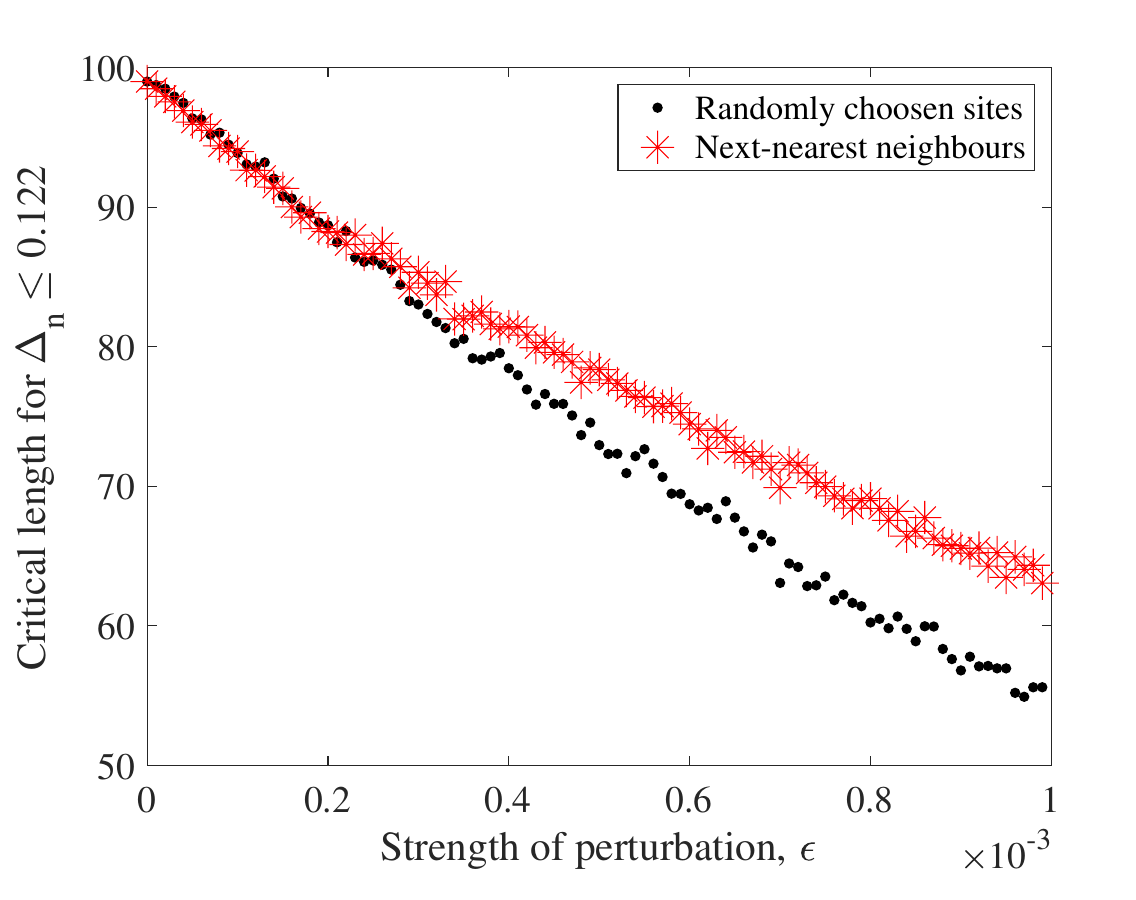} 
\par\end{centering}
\caption{The variation of the critical length $L_{C}$ with the strength of
perturbation $\epsilon$, up to a precision of $0.122$. The red star
correspond to the $L_{C}$ for next-nearest-neighbor interaction;
the black dots represent the cases where the sites that are interaction
are chosen at random. Each star/dot is an average of $1000$ random
instances of the Hamiltonian. The Hamiltonians are constructed with
 $c_{n}$ and $d_{n}$ randomly chosen from a uniform
distribution lying in $\left[0.95,1.05\right]$. }
\label{fig:P8} 
\end{figure}

Now, we look at a different setup where we randomly choose the adjacency structure as well as the interaction strengths. Figure~\ref{fig:P8} presents the behavior of critical length $L_{C}$ averaged over $1000$ random adjacency configurations. Each instance involves a different realization of the couplings along with randomly chosen set of interacting sites beyond nearest-neighbors. We observed from Fig.~\ref{fig:Pnew1} that the estimation algorithm performs worse in the presence of couplings beyond next-nearest-neighbor. Consistent with the error trends, the next-nearest-neighbor case shows higher average $L_{C}$, whereas systems with randomly placed interactions tend to result in shorter critical lengths.

These results suggest that systems with structured extended couplings -- such as next-nearest-neighbor interactions -- not only are more amenable to parameter estimation but also exhibit lower variance. In contrast, the presence of arbitrarily placed couplings introduces greater uncertainty and leads to higher average error, even when the total number and strength of interactions remain fixed.

\section{Conclusion}
\label{sec:con}
Our study sheds light on the importance of the type and arrangement of couplings. Investigating coupling strength holds practical importance in experimental settings, as it informs the selection of appropriate pulse types for experiments. We highlight the significance of the system's structural information. While the loss of physical information can be mitigated through modifications to the original scheme~\cite{Feng}, structural information plays a stronger and crucial role. A comprehensive understanding of the system's topology enables meaningful approximations and guides the adaptation of estimation schemes to realistic scenarios.

It is also conceivable that a similar extension could be achieved to unperturbed two- or three-dimensional scenarios. We have analyzed the estimation errors in a linear chain when perturbations effectively embed it into planar or spatial layouts. In the presence of perturbations, a network originally confined to two/three dimensions would likely introduce additional errors due to incomplete structural information. This is somewhat evident in Fig.~\ref{fig:Pnew1}, which demonstrates that the error is minimal when only next-nearest-neighbor couplings are present. However, when interactions occur between random pairs of sites, the error increases significantly.

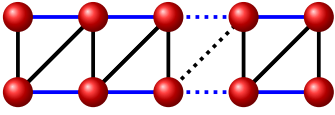
\begin{figure}[htbp]
\centering
\begin{tikzpicture}
        \foreach \i  in {1,...,2}{
            \foreach \j in {1,...,2,4}{
                \draw[line width=0.5mm,blue] (\j,\i) -- (\j+1,\i);
                \draw[line width=0.5mm,black] (\j,1) -- (\j+1,2);
            }    
        }
        \draw[line width=0.5mm,blue,dotted] (3,1) -- (4,1);
        \draw[line width=0.5mm,blue,dotted] (3,2) -- (4,2);
        \foreach \j in {1,...,5}{
            \draw[line width=0.5mm,black] (\j,1) -- (\j,2);
        }
        \draw[line width=0.5mm,black,dotted] (3,1) -- (4,2);
        
        \foreach \i  in {1,...,2}{
            \foreach \j in {1,...,5}{
                \shade[ball color=red] (\j, \i) circle (0.2);
            }    
        } 
\end{tikzpicture}
\caption{Topology corresponding to the Hamiltonian for system exhibiting nearest and next-nearest-neighbor interactions.
This is a two-dimensional representation of the topology shown in Fig.~\ref{fig:chain2}. The black and blue lines denote nearest and
next-nearest-neighbor couplings, respectively.}
\label{fig:chain3}
\end{figure}

It is worth observing that the one-dimensional topological structure associated with the couplings defined by the Hamiltonian featuring next-nearest-neighbor interactions (Fig.~\ref{fig:chain2}) can be equivalently represented as a two-dimensional system, as shown in Fig.~\ref{fig:chain3}. This transformation does not alter the underlying connectivity, but instead embeds the system in a higher-dimensional space where the interaction graph becomes locally simpler. 

\begin{figure}[htbp]
   \begin{minipage}{0.45\textwidth}
   \centering
    \begin{tikzpicture}
        \foreach \i  in {0,1}{
            \foreach \j in {0,1,...,6}{
                \draw[line width=0.5mm,black] (\j, \i) -- (\j+1,\i);
                }    
        }
        \draw[line width=0.6mm,white] (4, 1) -- (5,1);
        \draw[line width=0.5mm,black] (0, 0) -- (0,1);
        \draw[line width=0.5mm,black] (7, 0) -- (7,1);
        \foreach \i in {2,...,6}
        {
            \draw[line width=0.5mm,blue] (\i, 0) -- (\i,1);
        }
        \foreach \i  in {0,1}{
            \foreach \j in {0,1,...,7}{
                \shade[ball color=red] (\j, \i) circle (0.25);
            }    
        }
    \end{tikzpicture}
    \end{minipage}
    \\(a)\\
    \begin{minipage}{0.45\textwidth}
    \centering
    \begin{tikzpicture}
        \foreach \i  in {0,1}{
            \foreach \j in {0,1,...,6}{
                \draw[line width=0.5mm,black] (\j, \i) -- (\j+1,\i);
                } 
        }
        
        \draw[line width=0.5mm,black] (0, 0) -- (0,1);
        \draw[line width=0.5mm,black] (7, 0) -- (7,1);
        \foreach \i in {1,...,6}
        {
            \draw[line width=0.5mm,blue] (\i, 0) -- (\i,1);
        }
        \foreach \i  in {0,1}{
            \foreach \j in {0,1,...,7}{
                \shade[ball color=red] (\j, \i) circle (0.25);
            }    
        }
    \end{tikzpicture}
    \end{minipage}
    \\(b)
    \caption{Topology of qubits in IBM (a) QX3 and (b) QX5 architectures. Black and blue lines represent nearest and beyond nearest-neighbor interactions, respectively.}
    \label{fig:ibm1}
  \end{figure}
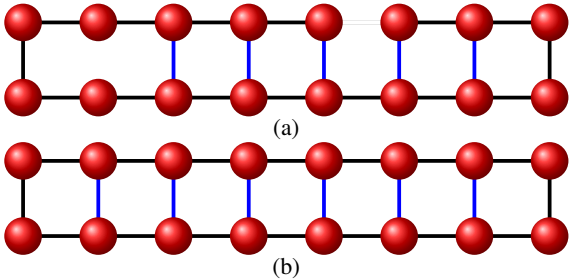

Such two-dimensional grid like structures closely resemble the physical layouts in experimental quantum devices, including those developed by IBM~\cite{IBM}. In Fig.~\ref{fig:ibm1}, we highlight two such IBM device topologies. Although our results are not tailored to a specific hardware, the structural similarities between our model and these architectures suggest a natural pathway for experimental validation. In particular, the estimation protocol -- which relies on spectral and phase information extracted from dynamical observables -- can be implemented in IBM's superconducting qubit devices. These platforms support programmable coupling maps and time-resolved measurements, which are essential to realize the phase reconstruction techniques discussed in Ref.~\cite{DB3}. Moreover, IBM's interface allows for fine-grained control over qubit interactions and readout, making them well suited for testing the sensitivity of our protocol to various coupling configurations.

By applying our methods to such architectures, one could investigate which interaction terms are critical for accurate Hamiltonian reconstruction and which can be neglected without significant loss of fidelity. This insight could help develop hardware-aware strategies for calibration, error mitigation and resource allocation in near-term quantum devices. 

\begin{acknowledgments}
This work was supported by the Australian Research Council (ARC) Centre of Excellence for Quantum Engineered Systems grant (CE 170100009).
\end{acknowledgments}

\appendix

\section{Couplings for nearest-neighbor interaction}
\label{app:A}
The tight-binding Hamiltonian in an $N$-dimensional Hilbert space
can be written as 
\[
    H_{C}=\sum_{n=1}^{N-1}c_{n}\left(|n\rangle\langle n+1|+|n+1\rangle\langle n|\right),
\]

where $\{|1\rangle,|2\rangle,\dots,|N\rangle\}$ are the basis vectors
and $c_{n}$'s are the nearest-neighbor couplings and $c_{n}>0$
(see Sec.~\ref{sec:frame}) 
. Let $\left\{ e_{k}\right\} $
be the eigenspectrum of $H_{C}$ and $\langle1|e_{k}\rangle$ be the
overlap of the $k$th eigenvector $\left|e_{k}\right\rangle $
with the reference state. We can see that 
\begin{align*}
&\langle1|H_{C}=c_{1}\langle2| \\
\implies&\langle1|H_{C}|e_{k}\rangle=e_{k}\langle1|e_{k}\rangle=c_{1}\langle2|e_{k}\rangle \\
\implies& c_{1}^{2}=\sum_{k=1}^{N}\left|e_{k}\langle1|e_{k}\rangle\right|^{2}.
\end{align*}
Thus, $c_{1}=\sqrt{\sum_{k=1}^{N}e_{k}^{2}\left|\langle1|e_{k}\rangle\right|^{2}}$
and $\langle2|e_{k}\rangle=\frac{e_{k}\langle1|e_{k}\rangle}{c_{1}}$.
Once we have $c_{1}$, we can recursively calculate the values of
$c_{n}$, $n=2,3,\dots,N-1$, as shown below:
\begin{align*}
&\langle n|H_{C}=c_{n}\langle n+1|+c_{n-1}\langle n-1| \\
\implies&\langle n|H_{C}|e_{k}\rangle=e_{k}\langle n|e_{k}\rangle=c_{n}\langle n+1|e_{k}\rangle+c_{n-1}\langle n-1|e_{k}\rangle \\
\implies&\langle n+1|e_{k}\rangle=\frac{e_{k}\langle n|e_{k}\rangle-c_{n-1}\langle n-1|e_{k}\rangle}{c_{n}}.
\end{align*}
Subsequently,
\begin{align*}
&c_{n}\langle n+1|e_{k}\rangle=e_{k}\langle n|e_{k}\rangle-c_{n-1}\langle n-1|e_{k}\rangle \\
\implies&\sum_{k=1}^{N}\left|c_{n}\langle n+1|e_{k}\rangle\right|^{2}=\sum_{k=1}^{N}\left|e_{k}\langle n|e_{k}\rangle-c_{n-1}\langle n-1|e_{k}\rangle\right|^{2} \\
\implies& c_{n}^{2}=\sum_{k=1}^{N}\left|e_{k}\langle n|e_{k}\rangle-c_{n-1}\langle n-1|e_{k}\rangle\right|^{2}.
\end{align*}
Therefore, we conclude that the knowledge of $\langle1|e_{k}\rangle$
and $e_{k}$ is sufficient to compute all the coupling constants.

\section{Error bounds for up to $n=2$}
\label{app:B}
We assume that we already have the values of $\left\{ e_{k}^{\epsilon}\right\} $
and $\left\{ \langle1|e_{k}^{\epsilon}\rangle\right\} $ and apply
the estimation scheme for nearest-neighbor interaction, but substitute $c_{n}\rightarrow c_{n}^{\epsilon}$
and $\langle n|e_{k}\rangle\rightarrow\langle n|e_{k}\rangle_{\varepsilon}$
to highlight the the possibility that the computed values may diverge
from the true values.

For $n=1$, $\left(c_{1}^{\epsilon}\right)^{2}=\sum_{k=1}^{N}\left(e_{k}^{\epsilon}\right)^{2}\left|\langle1|e_{k}^{\epsilon}\rangle\right|^{2}$
and $\langle2|e_{k}\rangle_{\varepsilon}\coloneqq{e_{k}^{\epsilon}\langle1|e_{k}^{\epsilon}\rangle}/{c_{1}^{\epsilon}}$.
The definition for $c_{1}^{\epsilon}$ does not take into account
the terms arising due to the perturbation, resulting in an error in
estimating the coupling $c_{1}$. Our objective is to find the relation
between $c_{1}^{\epsilon}$ and $c_{1}$, as a function of $\epsilon$.
Now,
\begin{align*}
&H^{\epsilon}|1\rangle=c_{1}|2\rangle+\epsilon d_{1}|3\rangle \\
\implies&\langle1|\left(H^{\epsilon}\right)^{2}|1\rangle=c_{1}^{2}+\epsilon^{2}d_{1}^{2} \\
\implies&\langle1|H^{\epsilon}\sum_{k=1}^{N}|e_{k}^{\epsilon}\rangle\langle e_{k}^{\epsilon}|H^{\epsilon}|1\rangle=c_{1}^{2}+\epsilon^{2}d_{1}^{2} \\
\implies&\sum_{k=1}^{N}\left(e_{k}^{\epsilon}\right)^{2}\left|\langle1|e_{k}^{\epsilon}\rangle\right|^{2}=c_{1}^{2}+\epsilon^{2}d_{1}^{2} \\
\implies&\left(c_{1}^{\epsilon}\right)^{2}=c_{1}^{2}+\epsilon^{2}d_{1}^{2}.
\end{align*}
Proceeding in a similar way, we arrive at an expression for $\left(c_{2}^{\epsilon}\right)^{2}$,
as shown below:

\begin{align*}
    \left(c_{2}^{\epsilon}\right)^{2} & =\sum_{k=1}^{N}\left|e_{k}^{\epsilon}\langle2|e_{k}\rangle_{\varepsilon}-c_{1}^{\epsilon}\langle1|e_{k}^{\epsilon}\rangle\right|^{2}
\end{align*}
\begin{align*}
    =\sum_{k=1}^{N}\left|e_{k}^{\epsilon}\frac{e_{k}^{\epsilon}\langle1|e_{k}^{\epsilon}\rangle}{c_{1}^{\epsilon}}-c_{1}^{\epsilon}\langle1|e_{k}^{\epsilon}\rangle\right|^{2}\
\end{align*}
\begin{align*}
    =\sum_{k=1}^{N}\left(e_{k}^{\epsilon}\right)^{2}\frac{\left(e_{k}^{\epsilon}\right)^{2}\left|\langle1|e_{k}^{\epsilon}\rangle\right|^{2}}{\left(c_{1}^{\epsilon}\right)^{2}}+\left(c_{1}^{\epsilon}\right)^{2}\sum_{k=1}^{N}\left|\langle1|e_{k}^{\epsilon}\rangle\right|^{2}
\end{align*}
\begin{align*}          -2c_{1}^{\epsilon}\sum_{k=1}^{N}e_{k}^{\epsilon}\frac{e_{k}^{\epsilon}\left|\langle1|e_{k}^{\epsilon}\rangle\right|^{2}}{c_{1}^{\epsilon}}
\end{align*}
\begin{align*}
    =\sum_{k=1}^{N}\frac{\left(e_{k}^{\epsilon}\right)^{4}\left|\langle1|e_{k}^{\epsilon}\rangle\right|^{2}}{\left(c_{1}^{\epsilon}\right)^{2}}+\left(c_{1}^{\epsilon}\right)^{2}-2\sum_{k=1}^{N}\left(e_{k}^{\epsilon}\right)^{2}\left|\langle1|e_{k}^{\epsilon}\rangle\right|^{2}.
\end{align*}

Setting $\left(c_{1}^{\epsilon}\right)^{2}=c_{1}^{2}+\epsilon^{2}d_{1}^{2}$
and using $\langle1|\left(H^{\epsilon}\right)^{4}|1\rangle=\sum_{k=1}^{N}\left(e_{k}^{\epsilon}\right)^{4}\left|\langle1|e_{k}^{\epsilon}\rangle\right|^{2}$,
the above equation reduces to
\[
\left(c_{2}^{\epsilon}\right)^{2}=\frac{\langle1|\left(H^{\epsilon}\right)^{4}|1\rangle}{\left(c_{1}^{\epsilon}\right)^{2}}-\left(c_{1}^{\epsilon}\right)^{2}.
\]
Expanding $\left(c_{1}^{\epsilon}\right)^{2}$ and ignoring $\mathcal{O}(\epsilon^{3})$,
we get
\[
\left(c_{2}^{\epsilon}\right)^{2}=c_{2}^{2}+\epsilon^{2}\left(d_{2}+\frac{d_{1}c_{3}}{c_{1}}\right)^{2}+\mathcal{O}(\epsilon^{3}).
\]

\bibliographystyle{unsrt}
\bibliography{ref}

\end{document}